\documentclass[12pt]{article}
\usepackage{bez123,calc,curves,ebezier,epic,eepic,graphicx,multiply,rotating}
\everymath{\displaystyle}
\usepackage{graphicx}
\usepackage{pst-all}
\usepackage{color}
\usepackage{amssymb}
\usepackage{amsmath}

\date{}
\usepackage{color}
\usepackage{blkarray}
\usepackage{calc}
\usepackage{tikz}

\newtheorem{prethm}{{\bf Theorem}}

\newenvironment{thm}{\begin{prethm}{\hspace{-0.5
				em}{\bf .}}}{\end{prethm}}
\newtheorem{prelemma}{{\bf Lemma}}

\newtheorem{preex}{{\bf Example}}

\newtheorem{preprop}{{\bf Proposition}}

\newenvironment{prop}{\begin{preprop}{\hspace{-0.5em}{\bf .}}}{\end{preprop}}
\newtheorem{precor}{{\bf Corollary}}

\newtheorem{preremark}{{\bf Remark}}

\newenvironment{remark}{\begin{preremark}{\hspace{-0.5
               em}{\bf.}}}{\end{preremark}}
\newtheorem{preprob}{{\bf Problem}}

\newtheorem{predefin}{{\bf Definition}}

\newenvironment{defin}{\begin{predefin}{\hspace{-0.5
               em}{\bf .}}}{\end{predefin}}
\newtheorem{preconj}{{\bf Conjecture}}

\newtheorem{preprobb}{{\bf Problem}}

\newtheorem{prelem}{{\bf Theorem}}

\newenvironment{proof}{{\bf Proof.}\rm }{\hfill{$\Box$}}

\newtheorem{presolution}{{\bf Solution.}}

\def\newpic#1{}

\title{\vspace{-2cm}\Large\bf A new vertex coloring heuristic and
corresponding chromatic number}

\author{\large\bf Manouchehr Zaker\footnote{mzaker@iasbs.ac.ir}
	\vspace{5mm}\\
	Department of Mathematics,\\
	Institute for Advanced Studies in Basic Sciences,\\
	Zanjan 45137-66731, Iran\\
}

\date{}
\begin{document}
\maketitle
\begin{abstract}
\noindent One method to obtain a proper vertex coloring of graphs using a reasonable number of colors is to start from any arbitrary proper coloring and then repeat some local re-coloring techniques to reduce the number of color classes. The Grundy (First-Fit) coloring and color-dominating colorings of graphs are two well-known such techniques. The color-dominating colorings are also known and commonly referred as {\rm b}-colorings. But these two topics have been studied separately in graph theory. We introduce a new coloring procedure which combines the strategies of these two techniques and satisfies an additional property. We first prove that the vertices of every graph $G$ can be effectively colored using color classes say $C_1, \ldots, C_k$
such that $(i)$ for any two colors $i$ and $j$ with $1\leq i< j \leq k$, any vertex of color $j$ is adjacent to a vertex of color $i$, $(ii)$ there exists a set $\{u_1, \ldots, u_k\}$ of vertices of $G$ such that $u_j\in C_j$ for any $j\in \{1, \ldots, k\}$ and $u_k$ is adjacent to $u_j$ for each $1\leq j \leq k$ with $j\not= k$, and $(iii)$ for each $i$ and $j$ with $i\not= j$, the vertex $u_j$ has a neighbor in $C_i$. This provides a new vertex coloring heuristic which improves both Grundy and color-dominating colorings. Denote by $z(G)$ the maximum number of colors used in any proper vertex coloring satisfying the above properties. The $z(G)$ quantifies the worst-case behavior of the heuristic. We prove the existence of $\{G_n\}_{n\geq 1}$ such that $\min \{\Gamma(G_n), b(G_n)\} \rightarrow \infty$ but $z(G_n)\leq 3$ for each $n$. For each positive integer $t$ we construct a family of finitely many colored graphs ${\mathcal{D}}_t$ satisfying the property that if $z(G)\geq t$ for a graph $G$ then $G$ contains an element from ${\mathcal{D}}_t$ as a colored subgraph. This provides an algorithmic method for proving numeric upper bounds for $z(G)$.
\end{abstract}




\section{Introduction}

\noindent All graphs in this paper are simple and undirected. Let $G$ be a graph with the vertex set $V(G)$. By a proper vertex coloring of $G$ we mean any function $C:V(G)\rightarrow \mathbb{N}$ such that for each adjacent vertices $u$ and $v$, $C(u)\not= C(v)$. In this paper by vertex colorings, we always mean proper vertex colorings of graphs. Denote by $\chi(G)$ the minimum number of colors used in any proper vertex coloring of a graph $G$. A very strong result proved by Zuckerman \cite{Zu} asserts that for every arbitrary and fixed $\epsilon>0$ if there exists a polynomial time algorithm ${\mathcal{A}}$ such that for every graph $G$ on $n$ vertices, ${\mathcal{A}}(G)/\chi(G) \leq n^{1-\epsilon}$ then $NP\subseteq P$, where ${\mathcal{A}}(G)$ denotes the number of colors used by the algorithm ${\mathcal{A}}$ to color the graph $G$. This in particular proves that it is a hard task to approximate the chromatic number of graphs within any constant factor. But graph coloring has wide applications in practical areas and hence exploring efficient graph coloring procedures is a necessary and vital goal. Efficient coloring heuristics is one of the important areas in graph theory and its algorithmic aspects (see e.g. \cite{CL,MT}). The greedy coloring procedure is one of the simplest heuristics which has been widely studied in the literature. The on-line version of greedy coloring scans the vertices according to an ordering of vertices (i.e. on-line presentation of the graph), assign the color $1$ to the first vertex and at each step color the current vertex by the smallest admissible color. In fact in any on-line coloring, the information of any input graph $G$ is presented gradually in on-line form i.e. vertex to vertex. A proper vertex coloring $C$ has Grundy property if for any two colors $i$ and $j$ with $i<j$, every vertex of color $j$ has a neighbor of color $i$. Any such coloring is called Grundy coloring. The Grundy coloring can be considered as the off-line version of the greedy coloring, in the sense that the entire information of the coloring is presented in its definition. Note that this off-line version could also be considered as a color reducing technique, as follows.

\noindent {\bf Name:} Grundy-type color reduction technique\\
\noindent {\bf Input:} A vertex coloring $C$ of $G$ with color classes $C_1, C_2, \ldots, C_t$\\
\noindent {\bf Output:} A vertex coloring satisfying the Grundy property using at most $t$ colors

\noindent 1. For $i=2$ up to $t$\\
2. ~~ Do for any vertex $v\in C_i$\\
3. ~~~~ If there exists $j<i$ such that $v$ has not any neighbor in $C_j$,\\
\hspace*{1.6cm} let $j(v)$ be the smallest such $j$\\
4. \hspace{1.1cm} Do $C_i \leftarrow C_i \setminus \{v\}$\\
\hspace*{1.6cm} $C_{j(v)} \leftarrow C_{j(v)} \cup \{v\}$ \\
5. ~~~~ If $C_i =\varnothing$\\
6. ~~~~~~~ Do remove $C_i$ from the lists\\
\hspace*{1.7cm} and for each $k>i$ rename $C_k$ by $C_{k-1}$\\
7. ~~~~ Else  $i\leftarrow i+1$\\
8. Return the refined color classes $C_1, C_2, \ldots$\\

\noindent It is easily seen that the output of the above technique is a proper coloring of $G$ satisfying the Grundy property. Note also that every greedy coloring procedure outputs a coloring satisfying Grundy property. Many of the coloring heuristics are special cases of greedy colorings in which an additional intelligent rule has been applied for choosing appropriate ordering of vertices. The greedy coloring obtained from the smallest-last-order, the so-called Max-Degree-Greedy and DSATUR \cite{B} are some examples of this sort. The other famous heuristic is termed Iterated Greedy in which the greedy coloring is applied repeatedly with different orderings of vertices. We refer the readers to \cite{C} and \cite{CL} for details and experimental properties of Iterated Greedy. All of these heuristics share a common property. The output of these heuristics are colorings satisfying the Grundy property. The other important point is that when these heuristics return output colorings (satisfying Grundy property) then no extra attempt is performed by the heuristics to reduce the number of colors in the output colorings. In this paper we introduce a new coloring method to reduce the color classes obtained from any greedy coloring of an input graph.

\noindent By a First-Fit (or Grundy) coloring of a graph $G$ we mean any proper vertex coloring of $G$ consisting of color classes $C_1, \ldots, C_t$ such that for each $i<j$ any vertex in $C_j$ has a neighbor in $C_i$. The First-Fit chromatic number (or Grundy number) denoted by $\chi_{FF}(G)$ (also by $\Gamma(G)$) is the maximum number of colors used in any First-Fit coloring of $G$. The First-Fit colorings have been widely studied in graph theory (see e.g. \cite{GL,Z2}). To determine the Grundy number is $NP$-complete even for complement of bipartite graphs \cite{Z1}. In some cases greedy coloring is an optimal coloring \cite{BMN}. The complexity aspects of Grundy number was studied in many papers. See e.g. \cite{BFKS,EGT,HS,Z2}. In this paper we also use the concept of {\it color-dominating coloring}. The color-dominating colorings are also known and commonly referred as {\rm b}-colorings.
Let $C$ be a proper vertex coloring of $G$ consisting of the color classes $C_1, \ldots, C_t$. Let $C_i$ be any color class in $C$. A vertex $v\in C_i$ is said to be a color-dominating vertex if for each $j\not= i$ there exists a neighbor of $v$ of color $j$ in $C$. Extracted from the concept of color-dominating coloring we introduce the following color-dominating technique to reduce the number of color classes. Let $C$ be any proper vertex coloring of $G$ consisting of the color classes $C_1, \ldots, C_t$. Then the color-dominating technique acts on the color classes of $C$ and applies the following method to reduce the number of classes. Take any class say $C_j$ from $C$. If $C_j$ contains a color-dominating vertex then we keep the class $C_j$ unchanged in the coloring. Otherwise, for any vertex $v$ of $C_j$ there exists a class $C_i$ such that $v$ has no neighbor in $C_i$. Remove $v$ from $C_j$ and transfer it to $C_i$. By this method the class $C_j$ becomes empty and is removed from the list of color classes. Repeat this technique for all other classes. We eventually obtain a coloring in which every color class contains a color-dominating vertex. The color-domination type technique is expressed in the following.

\noindent {\bf Name:} Color-domination type color reduction technique\\
\noindent {\bf Input:} A vertex coloring $C$ of $G$ with color classes $C_1, C_2, \ldots, C_t$\\
\noindent {\bf Output:} A color-dominating coloring using $p\leq t$ colors

\noindent 1. For $j=1$ up to $t$\\
2. ~ Do if $C_j$ has not any color-dominating vertex then resolve $C_j$ into
\newline {\hspace*{1.5cm} the other classes.}\\
3. ~~~~~~~ For $\ell=j+1$ up to $t$ rename the color class $C_{\ell}$ by $C_{\ell-1}$.\\
4. ~~~ Else the class $C_j$ remains unchanged, $j\leftarrow j+1$ and go to the step 2.\\
5. Go to the step 1 with the refined classes $C_1, C_2, \ldots$\\
6. Return $C_1, C_2, \ldots$\\

\noindent For convenient we call a color-dominating vertex (resp. color-dominating class) a CD vertex (resp. CD class). A coloring $C$ of $G$ is color-dominating (or {\rm b}-coloring) if each color class of $C$ is a color-dominating class. The idea of {\rm b}-coloring arises from the algorithmic approach to vertex coloring graphs. Let $G$ be a graph and $C_1, \ldots, C_k$ be any vertex coloring of $G$. For each $i$, if $C_i$ is not a color-dominating class then we can
transfer each vertex of $C_i$ to a suitable class in $C_1, \ldots, C_{i-1}, C_{i+1}, \ldots, C_k$. Hence by this method we can reduce the number of colors and eventually obtain a coloring which satisfies the {\rm b}-coloring property. The maximum number of colors used in any {\rm b}-coloring of $G$ is denoted by $b(G)$ (also by $\varphi(G)$) and is called the {\rm b}-chromatic number of $G$. The {\rm b}-chromatic number of graphs was firstly introduced and studied in \cite{IM}. Similar to Grundy coloring and number, the {\rm b}-coloring and {\rm b}-chromatic number of graphs have been the research subject of many articles. See the survey paper \cite{JP}.

\noindent For each positive integer $t$, a class of graphs denoted by ${\mathcal{A}}_t$ was constructed in \cite{Z2} which satisfies the following property. The Grundy number of any graph $G$ is at least $t$ if and only if $G$ contains an induced subgraph isomorphic to an element of ${\mathcal{A}}_t$. Any element of ${\mathcal{A}}_t$ is called $t$-atom. The complete graph on one vertex (resp. two vertices) are the only 1-atom (resp. 2-atom). The elements of ${\mathcal{A}}_{t+1}$ are constructed by the elements of ${\mathcal{A}}_t$. For each $t$ there exists only one unique tree $t$-atom denoted by $T_t$. We have $|V(T_t)|=2^{t-1}$ and that $T_t$ is the largest $t$-atom.
It was proved in \cite{Z2} that there exists a function $f(t)$ such that for every graph $G$ on $n$ vertices to determine weather $\Gamma(G)\geq t$ can be solved in at most $f(t)n^{2^{t-1}}$ time steps. In other words, to determine the Grundy number of graphs is a problem belonging to the complexity class $XP$. For definition of fixed-parameter related complexity classes such as $XP$, we refer the readers to the book \cite{C.et.al}. The $t$-atoms have been used and studied in some papers, e.g. \cite{BFKS, EGT}.
It was proved in \cite{HHB} that the Grundy number of trees can be determined in linear time. Let $T$ be any tree and $t$ any integer. In order to decide whether $\Gamma(T)\geq t$ we have to check if $T_t$ is subgraph of $T$ or not. For this purpose we can use the following result of Varma and Reyner in \cite{VR}. Let $T_1$ and $T_2$ be two arbitrary trees. There exists a polynomial time algorithm which determines whether $T_1$ is isomorphic to some subgraph of $T_2$. In other words, the Subtree Isomorphism is a polynomial time problem.

\noindent {\bf The aim and outline of the paper are as follow.} The aim of this paper is to introduce a new coloring heuristic which improves the greedy and color-dominating coloring heuristics. Because these coloring methods are commonly used in many color reducing algorithms, then they can be replaced by the new heuristic in order to achieve more optimal colorings. We show infinitely many graphs for which the new heuristic is intensively optimal than their greedy and color-dominating colorings. In the next section we prove that the vertices of any graph $G$ can be effectively colored satisfying three properties. This results in a new coloring heuristics and a new chromatic parameter denoted by $z(G)$. It follows that $z(G)\leq \min \{\Gamma(G), b(G)\}$. In Section 3, we construct a family of ``colored graphs" which we call $z$-atoms and prove that if $z(H)\geq t$ for some graph $H$ and integer $t$, then $H$ contains a $z$-atom $G$ as ``colored subgraph" such that $z(G)=t$. It implies that if $H$ does not contain any such $z$-atom then $z(H)< t$. This provides a polynomial time algorithm for proving upper bounds like $z(H)\leq t$, where $t$ is considered as a constant integer. In Section 4, we show that to determine $z(T)$ for trees $T$ is a polynomial time problem. The smallest tree $T$ with $z(T)=k$ has $(k-3)2^{k-1}+k+2$ vertices. For infinitely many trees the new heuristic behaves extremely better than the Grundy and color-dominating colorings.

\section{A new coloring heuristic}

\noindent In this section we first show that every graph admits a proper coloring which has simultaneously Grundy and color-dominating properties. In the previous section we presented the Grundy-type color reducing technique with details. This algorithm receives an arbitrary proper vertex coloring of $G$ and outputs a coloring satisfying the Grundy property. We summarize the algorithm in the following fact.

\begin{prop}
Let $C$ be any proper vertex coloring of a graph $G$. Then we can transform $C$ into a coloring which has Grundy property. Moreover, the transformation is done in at most $2|E(G)|$ steps.
\label{transform1}
\end{prop}

\noindent \begin{proof}
Let $C_1, \ldots, C_k$ be the color classes in $C$. The Grundy-type color reducing procedure transforms $C$ into the desired coloring of $G$. It's enough to determine the time complexity of the algorithm. Note that when a vertex $v$ is scanned and transmitted to a class say $C_j$ then it will remain in $C_j$ during the rest of the algorithm, because in this situation $v$ has a neighbor having color $i$, for each $1\leq i<j$. Also, we need at most $d_G(v)$ (the degree of $v$ in $G$) operations to specify the class $C_j$ corresponding to the vertex $v$. Hence, the whole process needs at most $2|E(G)|$ times steps.
\end{proof}

\noindent The next proposition achieves the first goal of this section.

\begin{prop}
Let $C$ be any Grundy coloring of a graph $G$. Then we can transform $C$ into a coloring which is simultaneously Grundy and color-dominating coloring. Moreover, the transformation is done in at most $2|E(G)|$ steps.
\label{transform2}
\end{prop}

\noindent \begin{proof}
\noindent Let $C_1, \ldots, C_k$ be the color classes in $C$. Any vertex in $C_k$ is adjacent to some vertex from any other class. Hence, $C_k$ contains color-dominating vertex (or vertices). Let $u$ be any vertex of $C_{k-1}$ which has a neighbor in $C_k$. Hence, $C_{k-1}$ contains color-dominating vertices too. We consider now $C_{k-2}$. If it contains a color-dominating vertex then we go to check the class $C_{k-3}$. Otherwise, corresponding to each vertex $v$ of $C_{k-2}$ there exists a smallest index $i(v)$ (in this case $i(v)\in \{k-1, k\}$) such that $v$ is not adjacent to any vertex in $C_{i(v)}$. Now, we move any vertex $v$ from $C_{k-2}$ to the class $C_{i(v)}$. Note that $C_{i(v)}$ is upper than $C_{k-2}$ and when a vertex is moved to an upper class then it will not be scanned again in the rest of procedure. After removing all vertices of $C_{k-2}$ then it becomes empty. We decrease by one unit the color of any vertex in $C_{k-1}\cup C_k$. Denote the resulting new coloring by $C'$. Note that $C'$ has Grundy property and uses $k-1$ colors. Also, the color classes $C'_{k-2}, C'_{k-1}$ in $C'$ contain color-dominating vertices. By continuing this method, assume that we have a Grundy coloring of $G$ consisting of the classes $D_1, \ldots, D_t$ such that each $D_i$ with $i\in \{t, t-1, \ldots, j+1\}$ contains at least one color-dominating vertex. We explain how to handle the class $D_j$. If $D_j$ has a CD vertex then we go to check the class $D_{j-1}$. Otherwise, for each vertex $v$ of $D_j$ there exists a smallest index $p(v)$ (in this case $p(v)\in \{j+1, \ldots, k\}$) such that $v$ is not adjacent to any vertex in $D_{p(v)}$. Now, we move any vertex $v$ from $D_j$ to the class $D_{p(v)}$. Then the class $D_j$ becomes empty. We decrease by one unit the color of any vertex in $D_{j+1}\cup \ldots \cup D_k$ and replace the coloring $D$ by the resulting new coloring $D'$. Note that $D'$ has Grundy property. The above procedure either eliminates a color class from the underlying coloring or finds a color-dominating vertex belonging to the same class. Since the elimination of classes can not occur more than $|V(G)|$ times then we eventually obtain a Grundy coloring which is also a color-dominating coloring. Also each vertex is scanned at most once in this procedure and for each vertex $v$ the number of steps to check the vertex $v$ is at most $d_G(v)$. Therefore, the whole procedure can be done in ${{\sum}_{v\in G}} d_G(v)=2|E(G)|$ time steps.
\end{proof}

\noindent We summarize the procedure explained in the proof of Proposition \ref{transform2} as the following heuristic. The proof shows that the heuristic returns a coloring which is Grundy and color-dominating.

\noindent {\bf Name:} Grundy Color-Dominating Heuristic
\newline {\bf Input:} A graph $G$ and an arbitrary Grundy coloring $C$ of $G$ with say $k$ colors
\newline {\bf Output:} A proper coloring $C'$ of $G$ which is simultaneously Grundy and color-dominating coloring, where at most $k$ colors are used

\noindent 1. $C_1, C_2, \ldots$ are color classes in a coloring $C$ satisfying Grundy property\\
\noindent 2. $k:=$ the number of classes in the coloring $C$\\
\noindent 3. For $j=k-2$ down to $2$\\
\noindent 4. ~~~Do if $C_j$ has a CD vertex then $j \leftarrow j-1$\\
\noindent 5. ~~~~~~~~Else for each $v\in C_j$,\\
\noindent 6. ~~~~~~~~~~~~~~~Do $C_{i(v)}\leftarrow C_{i(v)} \cup \{v\}$\\
\noindent 7. ~~~~~~~~~~~~For each $i>j$, $C_i \leftarrow C_{i-1}$\\
\noindent 8. ~~~Refine the color classes $C_1, C_2, \ldots$ and go to step $1$\\
\noindent 9. Return the final refined classes $C_1, C_2, \ldots$

\noindent Proposition \ref{transform2} shows that every graph admits a color-dominating coloring satisfying Grundy property. Let $G$ be a graph and $C$ be a Grundy and CD (i.e. color-dominating) coloring for $G$ using $k$ colors. There exist at least $k$ CD vertices with different colors in $C$. The subgraph of $G$ induced on these color-dominating vertices might be non-connected in general. This causes some problems in algorithmic analysis of such colorings, because in this case any $k$ vertices are potentially CD vertices and we have to explore all ${|V(G)| \choose k}$ subsets to obtain $k$ color-dominating vertices. In Theorem \ref{GCD}, we show that we can transform any Grundy and CD coloring $C$ of $G$ using $t$ colors to another Grundy and CD coloring $C'$ using say $k$ colors with $k\leq t$ and satisfying the additional property that $C'$ contains a set $D$ of color-dominating vertices with different colors such that $G[D]$ is isomorphic to a star graph i.e. $K_{1,k-1}$.

\noindent \begin{defin}
By a $z$-coloring of a graph $G$ we mean any proper vertex coloring $C$ of $G$ using say $k$ colors which is simultaneously Grundy and color-dominating coloring and contains a set $\{u_1, \ldots, u_k\}$ of color-dominating vertices such that for each $j$, the color of $u_j$ is $j$ and for each $j\not=k$, $u_k$ is adjacent to $u_j$. Denote by $z(G)$ ($z$-number of $G$) the maximum number of colors used in any $z$-coloring of $G$.\label{zcoloring}
\end{defin}

\noindent In Theorem \ref{GCD}, we prove that every graph $G$ admits a $z$-coloring. In order to prove the theorem we need the concept of nice vertex.

\begin{defin}
Let $C$ be any proper coloring of $G$ using $t$ colors. A vertex $v$ is called a {\it nice vertex} for the coloring $C$ if the color of $v$ is $t$ in $C$ and $v$ is adjacent to at least $t-1$ color-dominating vertices with $t-1$ different colors in $C$.
\end{defin}

\begin{thm}
Let $G$ be a graph on $n$ vertices and $m$ edges. Let $C$ be any Grundy and color-dominating coloring for a graph $G$ using $t$ colors. There exists an ${\mathcal{O}}(mn)$ procedure which transforms $C$ into a Grundy and CD coloring $C'$ using say $k$ colors such that $k\leq t$ and $C'$ contains a set $\{u_1, \ldots, u_k\}$ of color-dominating vertices such that for each $j$ the color of $u_j$ is $j$ (in $C'$) and $u_k$ is adjacent to $u_j$ for each $j\not=k$.\label{GCD}
\end{thm}

\noindent \begin{proof}
Let $C_t$ be the class of vertices with color $t$ in $C$. Since $C$ is Grundy then each vertex in $C_t$ is a CD vertex. If the class $C_t$ contains a nice vertex then $C$ satisfies the conditions of proposition. Otherwise, corresponding to each vertex $u\in C_t$ there exists a minimum color $i(u)$ such that $u$ is not adjacent to any CD vertex of color $i(u)$. It follows that for each vertex $u\in C_t$ and for each vertex $w\in N(u)$ of color $i(u)$ in $C$, there exists a smallest color $j(w)$ such that $j(w)$ is not appeared in the neighborhood of $w$. We have $i(u)< j(w) < t$. Note that to check if $u$ is nice vertex is done in ${\mathcal{O}}(m)$ time steps. During this check and in case that $u$ is not nice vertex the values $i(u)$ and $j(w)$ are also determined for each $w\in N(u)$. Let now $u_0$ be an arbitrary vertex of $C_t$. We make a local recoloring as follows. Change the color of $u_0$ from $t$ to $i(u_0)$ and assign the color $j(w)$ to any neighbor $w$ of $u_0$ whose color in $C$ is $i(u_0)$. Observe that after this recoloring the vertices $u_0$ and its neighbors remain Grundy vertices. If necessary using the techniques of Propositions \ref{transform1} and \ref{transform2} modify the color of some vertices in $C$ and obtain a Grundy and CD coloring $C'$. Note that either there exists no class of color $t$ in $C'$ or the number of vertices of color $t$ in $C'$ is strictly less than $|C_t|$. We repeat this technique to the other vertices of color $t$ in $C'$. Such possible vertices of color $t$ in $C'$ are already of color $t$ in the coloring $C$. Either we obtain a nice vertex in the underlying coloring and therefore the procedure is finished at this step or we obtain a Grundy and CD coloring $C''$ which uses less than $t$ colors. Again, if $C''$ has a nice vertex then $C''$ satisfies the conditions of the proposition and proof completes. Otherwise, let $C''_p$ be the class of colors with the maximum color in $C''$. Instead of $C$ and $C_t$ we repeat the same technique for $C''$ and $C''_p$. By continuing this method either we obtain a coloring with a nice vertex or one color class is removed and we obtain a Grundy and CD coloring with fewer colors. Obviously the color classes can not be removed more than $|V(G)|=n$ times. We conclude that eventually a coloring $C'''$ satisfying the desired conditions is achieved.

\noindent Obviously, the number of total iterations is at most $n$. By Propositions \ref{transform1} and \ref{transform2}, each iteration needs time complexity ${\mathcal{O}}(m)$. As explained before, to check if an individual vertex is nice takes ${\mathcal{O}}(m)$ steps. Hence the entire process of exploring nice vertices is done in ${\mathcal{O}}(mn)$ time steps. Overall, the transformation of $C$ into $C'''$ is performed in ${\mathcal{O}}(mn)$ time steps. This completes the proof.
\end{proof}

\noindent In the light of Proposition \ref{transform2} and Theorem \ref{GCD} we introduce a new coloring heuristic and an associated chromatic parameter. By the $z$-coloring heuristic we mean the coloring heuristic explained as follows. By the proof of Theorem \ref{GCD}, the following heuristic outputs in ${\mathcal{O}}(mn)$ time steps a $z$-coloring on any graph with $n$ vertices and $m$ edges.

\noindent {\bf Name:} $z$-coloring heuristic\\
\noindent {\bf Input:} A graph $G$\\
\noindent {\bf Output:} A $z$-coloring of $G$ satisfying Definition \ref{zcoloring}

\noindent 1. Start from any proper vertex coloring $C$ of $G$\\
\noindent 2. Apply the Grundy technique and Grundy Color-Dominating heuristic for $C$ and obtain a Grundy and color-dominating coloring consisting of the color classes $C_1, C_2, \ldots$\\
\noindent 3. $t:=$ the number of colors in $C$\\
\noindent 4. ~~For each vertex $u\in C_t$\\
\noindent 5. ~~~~~Do if $u$ is nice vertex then go to step 11 and return $C$\\
\noindent 6. ~~~~~~~~~~Else, determine a minimum color $i(u)$ such that $u$ is not adjacent\\
\hspace*{2.7cm} to any CD vertex of color $i(u)$\\
\noindent 7. ~~~~~~~~~~~~~~~~~For each vertex $w\in N(u)$ of color $i(u)$ in $C$\\
\noindent 8. ~~~~~~~~~~~~~~~~~~~~~~Do determine a smallest color $j(w)$ such that $j(w)$ is not\\
\hspace*{4cm} appeared in the neighborhood of $w$\\
\noindent 9. ~~~~~~~~~~~~~~~~~~~~~~Do $c(u) \leftarrow i(u)$ and $c(w) \leftarrow j(w)$\\
\noindent 10. ~~~~~~~~~~~~~~~~~~~~~Refine the color classes and go to the step 2\\
\noindent 11. Return the final classes $C_1, C_2, \ldots$

\noindent We have the following proposition.

\begin{prop}
For any graph $G$, $z(G)\leq \min \{\Gamma(G), b(G)\}$. Moreover, there exists an infinite sequence of graphs $\{G_t\}_{t=1}^{\infty}$ such that $\Gamma(G_t)\rightarrow \infty$ and $b(G_t)\rightarrow \infty$ as $t\rightarrow \infty$ but for each $t$, $z(G_t)=3$.\label{infinityexample}
\end{prop}

\begin{figure}
	\hspace*{2cm}\includegraphics[width=10.6cm,height=3.3cm]{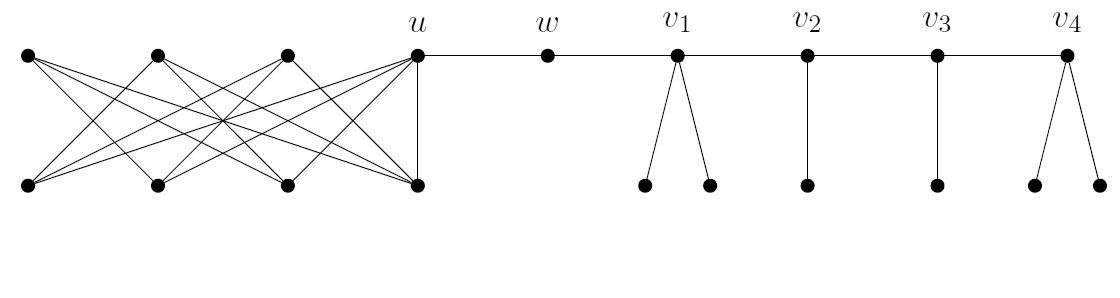}
	\caption{The graph $G_t$ in Proposition \ref{infinityexample} for $t=4$}\label{Gt}
\end{figure}

\noindent \begin{proof}
The inequality clearly holds. For each positive integer $t$, define $H_t=K_{t, t}\setminus (t-1)K_2$. It is easily seen that $\Gamma(H_t)=t+1$. It is also observed that $H_t$ does not admit any {\rm b}-coloring using more than 2 colors. Construct another sequence of graphs as follows. Let $t$ be any arbitrary and fixed positive integer. First consider a path on $t$ vertices on the vertex set $\{v_1, \ldots, v_{t}\}$. Then attach $t-2$ leaves to $v_1$ and additional $t-2$ leaves to $v_{t}$. Also, for each $i$ with $2\leq i \leq t-1$, attach $t-3$ leaves to $v_i$. All of these leaves are distinct. Denote the resulting graph by $F_t$. We claim that $b(F_t)=t$. For each $i$, assign color $i$ to the vertex $v_i$. Now, using the leaf vertices we can extend this pre-coloring to a {\rm b}-coloring of $F_t$ using $t$ colors. Since $\Delta(F_t)=t-1$ and $b(F_t)\leq \Delta(F_t)+1$ then $b(F_t)=t$. Now, consider the graph $H_t$ in one side and the graph $F_t$ in other side. Let $u$ be any vertex of degree $t$ in $H_t$. Put one edge between the vertex $u$ from $H_t$ and the vertex $v_1$ from $F_t$ and then subdivide this edge by putting an additional vertex say $w$. Denote the resulting graph by $G_t$, where there exists a path of length two between $u$ and $v_1$. The graph $G_4$ is depicted in Figure \ref{Gt}. Obviously, $\Gamma(G_t)\rightarrow \infty$ and $b(G_t)\rightarrow \infty$ as $t\rightarrow \infty$. To complete the proof we argue that $z(G_t)\leq 3$ for each $t$. Its proof is simple. Assume that $G_t$ admits a $z$-coloring using four or more colors. Let $v$ be a vertex of color $4$ in a $z$-coloring of $G_t$. The vertex $v$ needs at least three neighbors of degree at least three. Hence $v\not\in V(F_t)$. Also $v$ can not be $w$ because the degree of $w$ in $G_t$ is two. Therefore $v$ should be a vertex in $H_t$. As we said $v$ needs three neighbors of degree at least three. None of such neighbors of $v$ can be $w$ because $w$ has degree two. It follows that every $z$-coloring of $G_t$ using more than three colors reduces to a {\rm b}-coloring with more than three colors in $H_t$. But as explained before this is not possible.
\end{proof}

\noindent A graph parameter $p$ is said to be monotone if for every graph $G$ and any induced subgraph $H$ of $G$, $p(H)\leq p(G)$. An interesting property of $z$-coloring is that it's not monotone, i.e. if $H$ is an induced subgraph in $G$ then it does not necessarily imply that $z(H)\leq z(G)$. For example let $G=K_{t+1, t+1}\setminus tK_2$ and $H=K_{t, t}\setminus tK_2$. Then $H\subseteq G$, $z(G)=2$ but $z(H)=t$. This is interesting because suppose that a graph $G$ has a large $z(G)$ and let $C$ be a $z$-coloring of $G$ using $z(G)$ colors. This does not provide a good upper bound for the ordinary chromatic number of $G$, but there is still a chance that if you add one vertex $u$ (and in some cases two adjacent vertices $u, u'$) to $G$ and $u$ (resp. $u, u'$) becomes a color-dominating vertex for $C$ then the $z$-number reduces significantly and therefore the graph properly colored with an smaller number of colors. We present some examples.
Let $P_5$ be the path on five vertices $v_1, \ldots, v_5$, where $v_1, v_5$ are its leaves and $v_3$ its central vertex. Assign color 3 to $v_3$, color 2 to $v_2, v_5$, and assign color 1 to $v_1, v_4$. It follows that $z(P_5)=3$. Add a new vertex $u$ and put edges between $u$ and $v_1, v_3, v_5$. Assign color 4 to $u$ and obtain a Grundy coloring of the new graph using 4 colors. Apply the technique of Proposition \ref{transform2} to these data and obtain a 2-coloring for the whole graph and hence for the $P_5$. The second example is the cycle on 6 vertices $C_6$. Color its vertices consecutively $3, 2, 1, 3, 2, 1$ and obtain a $z$-coloring using 3 colors. A main difference between these two examples is that in the case of $C_6$ all vertices are color-dominating. In this situation we add two adjacent vertices $u, u'$ to $C_6$ and put enough edges between $u, u'$ and $C_6$ such that they become color-dominating and share no common neighbor. Assign colors 4, 5 to $u, u'$, respectively. Denote the resulting graph by $H$ and coloring by $C$. The coloring $C$ is Grundy coloring using 5 colors. Apply the technique of Proposition \ref{transform2} for $(H,C)$. If $H$ is bipartite then we obtain a proper 2-coloring for $H$ and hence for $C_6$. Otherwise, we obtain a 3-coloring for $C_6$. Note that the graphs $G=K_{t+1, t+1}\setminus tK_2$ and $H=K_{t, t}\setminus tK_2$ satisfying $z(G)=2$ but $z(H)=t$ (mentioned in the proof of Proposition \ref{infinityexample} and at the beginning of this paragraph) have properties similar to our second example. In fact the graph $G$ is systematically obtained from $H$ using this technique, i.e. two adjacent vertices are added to $H$ and so on.

\noindent Based on the preceding examples and comments we add an important extra step to our proposed heuristic to achieve a complementary from of the heuristic.

\noindent {\bf The complementary form of the heuristic:}

\noindent Assume that the $z$-coloring heuristic has colored a given graph $G$ and output the color classes $C_1, \ldots, C_t$. Denote the Cartesian product of these color classes by $C_1 \times C_2 \times \ldots \times C_t$. For any element $(v_1, v_2, \ldots, v_t)\in C_1 \times C_2 \times \ldots \times C_t$, define a graph denoted by $G_{(v_1, v_2, \ldots, v_t)}$ as follows.
Corresponding to $(v_1, v_2, \ldots, v_t)$ add a new vertex to $G$ and connect the new vertex to the vertex $v_i$, for each $i=1, \ldots, t$. For each element $(v_1, v_2, \ldots, v_t)$, apply the $z$-coloring heuristic to $G_{(v_1, v_2, \ldots, v_t)}$. Each resulting coloring comprises a vertex coloring of $G$. Choose the one with minimum number of colors.

\section{Construction of $z$-atoms and their properties}

\noindent In order to explain what we accomplish in this section, we need to introduce a concept.

\begin{defin}
Let $H$ and $G$ be two graphs. Let $C$ be a proper vertex coloring of $H$. For each vertex $v$ of $H$, denote the color of $v$ in $H$ by $C(v)$. We say $(H,C)$ is embedded in $G$ if there exists an injective function $f:V(H)\rightarrow V(G)$ such that the following two conditions hold:

\noindent (i) for each two vertices $u, v$ of $H$ if $uv\in E(H)$ then $f(u)$ and $f(v)$ are adjacent in $E(G)$ (i.e. $H$ is isomorphic to a subgraph of $G$)

\noindent (ii) if $C(u)=C(v)$ (i.e. $u, v$ have identical colors in $H$) then $f(u)$ and $f(v)$ are not adjacent in $E(G)$.
\end{defin}

\noindent For example, let $H$ be the path on four vertices and $C$ be a Grundy coloring of $H$ using three colors. Let $G$ be the cycle on four vertices. Then $H$ is subgraph of $G$ but $(H,C)$ is not embedded in $G$.

\noindent In the following for each positive integer $t$ we construct a collection of graphs ${\mathcal{D}}_t$ such that each member of ${\mathcal{D}}_t$ is a colored graph such as $(G,C)$, where $C$ is a $z$-coloring of $G$ using $t$ colors. The coloring $C$ is called the canonic $z$-coloring of $G$. We prove that ${\mathcal{D}}_t$ has the following property. Let $L$ be any graph such that $z(L)\geq t$. Then there exists $(G,C)$ in ${\mathcal{D}}_t$ such that $(G,C)$ is embedded in $L$. As we explained in the paragraph before Section 3, we are lucky that the converse of this result does not necessarily hold. In the following we say a graph $H$ is edge-minimal with respect to $z$-number (or simply edge-minimal) if for any edge $e$ of $H$ we have $z(H\setminus e)\leq z(H)-1$. Let $S_{1,t}$ be the star graph consisting of a central vertex of degree $t$ and $t$ leaves $u_1, u_2, \ldots, u_t$ adjacent to the central vertex. We assign the color $t+1$ to the central vertex and the color $i$ to the vertex $u_i$, for each $i\in \{1, 2, \ldots, t\}$. Let $t\geq 0$ be any integer. We construct the elements of ${\mathcal{D}}_{t+1}$ during two distinct phases. Starting from the very star graph $S_{1,t}$, in Phase I we generate all possible graphs $H$ together with a proper vertex coloring $C$ for $H$ satisfying the following properties

\begin{itemize}
\item{The graph $H$ contains $S_{1,t}$ as subgraph and for any $p\in \{1,2, \ldots, t\}$, the color of $u_p$ in $C$ is $p$. Moreover, for any $p$ with $1\leq p \leq t-1$ and for any $\ell$ with $p+1\leq \ell \leq t$, the vertex $u_p$ has a neighbor of color $\ell$ in $H$.~~~~~~~{\bf $\clubsuit$}}
\item{For any edge $e$ of $H$, $H\setminus e$ does not satisfy {\bf $\clubsuit$}.}
\end{itemize}

\noindent All colored graphs $(H,C)$ produced in Phase I, enter Phase II. In Phase II a graph $(H,C)$ is extended to a collection of colored graphs each of which is typically denoted by $(G,C')$ such that $(H,C)$ is embedded in $G$ and the restriction of $C'$ to the vertices of $H$ is identical to $C$. Moreover, $C'$ is a $z$-coloring of $G$ using $t+1$ colors and $G$ is edge-minimal graph with respect to $z$-number, where the vertices of $S_{1,t}$ are the color-dominating vertices with colors $t+1, t, \ldots, 2, 1$. Finally, the collection ${\mathcal{D}}_{t+1}$ is defined as the collection of all such graphs $(G,C')$ produced in Phase II. In fact, the Phase II produces all edge-minimal colored graphs with $z$-number $t+1$ and containing $S_{1,t}$. In terms of graph notations, the following sequence denotes the evolvement of our constructions. $$S_{1,t} \longrightarrow (H,C) \longrightarrow (G,C').$$
\noindent In Theorem \ref{z-atom}, we prove that for every graph $L$ if $z(L)\geq t$ then there exists a member $(G,C')$ from ${\mathcal{D}}_t$ such that $(G,C')$ is embedded in $L$.

\noindent {\bf Construction Phase I:}

\noindent We consider the star graph $S_{1,t}$ with leaf vertices $u_1, u_2, \ldots, u_t$ and its coloring using $t+1$ colors and want to generate all possible edge-minimal graphs $(H,C)$ satisfying the above-mentioned conditions {\bf $\clubsuit$}. In fact, for any $p$ and $\ell\geq p+1$, the vertices $u_1, \ldots, u_p$ need neighbors (in $H$) with color $\ell$ (in $C$). These neighbors are not necessarily distinct. It follows that, for each $j$ with $2\leq j \leq t$ we should have $j-1$ (not necessarily distinct) neighbors of color $j$ in $(H,C)$. Hence, we need potentially say $m_j$ distinct vertices of color $j$ in $(H,C)$, where $m_j$ can be any value with $1 \leq m_j \leq j-1$. Note that we have already a vertex of color $j$ in the star graph, i.e. the vertex $u_j$. This means that one possibility is that all vertices $u_1, \ldots, u_{j-1}$ are adjacent to $u_j$ in order to have a neighbor of color $j$. But this is only one of the possibilities. Another important point is that if we satisfy the above-mentioned conditions for the neighbors of $u_1, u_2, \ldots, u_t$ then the task is finished since because of minimality we need no extra vertices or edges. Based on this background, we explain Phase I gradually as follows:

\noindent For any $j$, $2\leq j\leq t$, let $m_j$ be an arbitrary integer with $0\leq m_j\leq j-1$. If $m_j=0$ then define $M_{j,0}=\{u_j\}$. If $m_j\geq 1$ then let $M_{j,m_j}$ be a set consisting of $u_j$ and $m_j$ additional vertices. We have $|M_{j,m_j}|=m_j+1$ and for any $j, j'$ with $j\not= j'$, $M_{j,m_j}\cap M_{j',m_{j'}}=\varnothing$.

\noindent Corresponding to any fixed selection of $m_2, m_3, \ldots, m_t$ and any set of surjective functions $\{f_{j,m_j}:j=2, 3, \ldots, t\}$, where $f_{j,m_j}: \{1, 2, \ldots, j-1\}\rightarrow M_{j,m_j}$ define a graph $H$ as follows.

\noindent Add the vertices of $M_{2,m_2}\cup M_{3,m_3} \cup \ldots \cup M_{t,m_t}$ as all distinct and extra vertices to $S_{1,t}$. Note that if $m_j=0$ for some $j$, then no vertex corresponding to $j$ is added to the graph $S_{1,t}$ because as specified before, we have $M_{j,0}=\{u_j\}$ and $u_j$ is already present in $S_{1,t}$ (and therefore in $H$). Now, for each $j\in \{2, \ldots, t\}$ and any value $p$, $1\leq p\leq j-1$, put an edge between $u_p$ and $f_{j,m_j}(p)$. Assign the color $j$ to all vertices in $M_{j,m_j}$.

\noindent Denote the resulting graph and vertex coloring by $H$ and $C$, respectively. Observe that for any $p$ and any $\ell\in \{p+1, \ldots, t\}$, the vertex $u_p$ has a neighbor of color $\ell$ in $H$. Because, $u_p$ is adjacent to $f_{\ell,m_{\ell}}(p)$ and the color of $f_{\ell,m_{\ell}}(p)$ is $\ell$ in the coloring $C$ (in fact $f_{\ell,m_{\ell}}(p) \in M_{\ell,m_{\ell}}$).

\noindent Before we proceed toward the second phase of construction we prove a result concerning the collection of graphs generated in Phase I.

\begin{prop}
Let $G$ be any graph and $C$ be a $z$-coloring of $G$ using $t+1$ colors. Then there exist $t+1$ vertices $u_1, u_2, \ldots, u_{t+1}$ of colors $1, 2, \ldots, t+1$ in $C$, respectively and a colored graph $(H,C')$ satisfying the following properties.\\
\noindent (i) For any $i$, $u_{t+1}$ is adjacent to $u_i$.\\
\noindent (ii) For each $i$, $1\leq i\leq t+1$, $u_i$ is a color-dominating vertex of color $i$.\\
\noindent (iii) Started from the star graph consisting of the vertices $u_1, u_2, \ldots, u_{t+1}$, the graph $(H,C')$ is subgraph of $G$ and one of the graphs generated in Phase I, where the coloring $C'$ is the restriction of $C$ to the vertices of $H$.\label{phaseI}
\end{prop}

\noindent \begin{proof}
Since $C$ is a $z$-coloring then by the definition there exist $t+1$ color-dominating vertices say $u_1, u_2, \ldots, u_{t+1}$ in $G$ which satisfy the conditions $(i)$ and $(ii)$. Since $u_1, u_2, \ldots, u_{t+1}$ are color-dominating vertices then corresponding to any value $j$ with $2\leq j \leq t$ there exists a set
$M_j$ of vertices having color $j$ in $C$ such that for any $p$ with $1\leq p\leq j-1$ the vertex $u_p$ is adjacent to some vertex in $M_j$. The set $M_j$ can be easily chosen so that each vertex of which has a neighbor among $u_1, \ldots, u_{j-1}$. Note that the vertex $u_j$ itself may or may not belong to $M_j$. We consider the subgraph $H$ of $G$ on the vertex set $\{u_1, u_2, \ldots, u_{t+1}\} \cup M_2 \cup M_3 \cup \cdots \cup M_t$ and consisting of the above-mentioned edges between $\{u_1, u_2, \ldots, u_{t+1}\}$ and $M_2 \cup M_3 \cup \cdots \cup M_t$ as well as the edges among $u_1, \ldots, u_{t+1}$. Let $C'$ be the coloring of $H$ obtained from the restriction of $C$ to the vertices of $H$. Considering the sets $M_j$ and functions which map any vertex from $\{u_1, \ldots, u_{j-1}\}$ to its neighbor in $M_j$, we conclude that $(H,C')$ is one of the graphs produced in Phase I. This completes the proof.
\end{proof}

\noindent {\bf Construction Phase II:}

\noindent We need first to define the concept of Grundy class. Let $H$ be an arbitrary graph and $C$ a proper vertex coloring of $H$ consisting of the color classes $C_1, \ldots, C_t$. A color class say $C_j$ is said to be Grundy class if for any color $i<j$ any vertex $u$ of $C_j$ has a neighbor of color $i$ in $H$. In the following we introduce the operation ${\mathcal{G}}_k$, where $k$ is an arbitrary natural number. The operation ${\mathcal{G}}_k$ operates on arbitrary colored graphs $(H,C)$ and outputs a collection of colored supergraphs of $H$. Let the color classes in $(H,C)$ be $C_1, \ldots, C_k, \ldots, C_t$. There are two possibilities concerning $C_k$. If the class $C_k$ is Grundy then ${\mathcal{G}}_k$ operates on $(H,C)$ and outputs the same graph, i.e. ${\mathcal{G}}_k(H,C)=(H,C)$ in this case. If $C_k$ is not a Grundy class then the output of ${\mathcal{G}}_k$ is a collection of colored graphs typically denoted by $(G,C')$ such that the following hold. The graph $H$ is an induced subgraph of $G$, the restriction of $C'$ to the vertices of $H$ is the same as the coloring $C$, both colorings $C$ and $C'$ have the same number of colors and the $k$-th class in $C'$ is a Grundy class.
We may interpret that the operation ${\mathcal{G}}_k$ ``Grundify" the color class $k$ in its input graphs $(H,C)$.

\noindent {\bf The operation ${\mathcal{G}}_k$:}

\noindent Let $(H,C)$ be an arbitrary input graph for the operation ${\mathcal{G}}_k$, where $C_k$ is not a Grundy class in $(H,C)$. Let $i\in \{1, \ldots, k-1\}$ be an arbitrary and fixed integer and let $C_k^i$ be a subset of $C_k$ consisting of the vertices with no neighbor of color $i$ in $C$.
Recall that $C_i$ is the set of all vertices having color $i$ in $C$. Denote an arbitrary subset of $C_k^i$ by $S_i$. Let $S_i$ be an arbitrary and fixed subset of $C_k^i$. Let $f_i$ be any arbitrary and fixed function from $S_i$ to $C_i$. Let also $m_i$ be any arbitrary value with $1\leq m_i \leq |C_k\setminus S_i|$. Consider some extra vertices $w^i_1, \ldots, w^i_{m_i}$ and let $g_i$ be any arbitrary surjective (onto) function from $C_k^i\setminus S_i$ onto $\{w^i_1, \ldots, w^i_{m_i}\}$. Set ${\mathcal{S}}=(S_1, \ldots, S_{k-1})$, $\mathfrak{f}=(f_1, \ldots, f_{k-1})$, $\mathfrak{m}=(m_1, \ldots, m_{k-1})$ and $\mathfrak{g}=(g_1, \ldots, g_{k-1})$. Now corresponding to $({\mathcal{S}}, \mathfrak{f}, \mathfrak{m}$, $\mathfrak{g})$ we construct a graph $H_{\mathcal{S},\mathfrak{f},\mathfrak{m}, \mathfrak{g}}$ as follows.

\noindent Recall that corresponding to each $i$, $1\leq i \leq k-1$ we have taken a subset $S_i\subseteq C_k^i$, a function $f_i:S_i\rightarrow C_i$, an integer $m_i$ with $1\leq m_i \leq |C_k\setminus S_i|$ and finally a surjective function $g_i: C_k^i\setminus S_i \rightarrow \{w^i_1, \ldots, w^i_{m_i}\}$. Now, corresponding to each $i$ we perform the following operations. Add the $m_i$ extra vertices $w^i_1, \ldots, w^i_{m_i}$ to the graph $H$ and put an edge between any vertex $u\in C_k^i\setminus S_i$ and $g_i(u)$. Note that $g_i(u)\in \{w_1, \ldots, w_{m_i}\}$. Next, we put an edge between any vertex $u'\in S_i$ and $f_i(u')$. Note that $f_i(u')\in C_i$. Denote the resulting graph by $H_{\mathcal{S},\mathfrak{f},\mathfrak{m}, \mathfrak{g}}$. We have $$V(H_{\mathcal{S},\mathfrak{f},\mathfrak{m}, \mathfrak{g}})=V(H)\cup (\bigcup_{i=1}^{k-1} \{w^i_1, \ldots, w^i_{m_i}\}).$$
\noindent Define a proper vertex coloring $C'$ of $H_{\mathcal{S},\mathfrak{f},\mathfrak{m}, \mathfrak{g}}$ as follows. For any vertex $v\in H$ set $C'(v)=C(v)$ and for each $j\in \{1, \ldots, m_i\}$, $C'(w^i_j)=i$. Note that in $H_{\mathcal{S},\mathfrak{f},\mathfrak{m}, \mathfrak{g}}$ the class of vertices of color $k$ is a Grundy class.

\begin{remark}
Let $(H,C)$ be any typical graph constructed in Phase I. Let $C_k$ be a color class in $(H,C)$ which is not a Grundy class. Then ${\mathcal{G}}_k$ operates on $(H,C)$ and generates a family of graphs of the form $H_{\mathcal{S},\mathfrak{f},\mathfrak{m}, \mathfrak{g}}$. In case that $C_k$ is Grundy class then ${\mathcal{G}}_k$ leaves $(H,C)$ unchanged. In any case the class of vertices of color $k$ in each member of ${\mathcal{G}}_k (H,C)$ is Grundy class.\label{grundify}
\end{remark}

\begin{figure}
	\hspace*{-4cm}
\includegraphics[width=25cm,height=20cm]{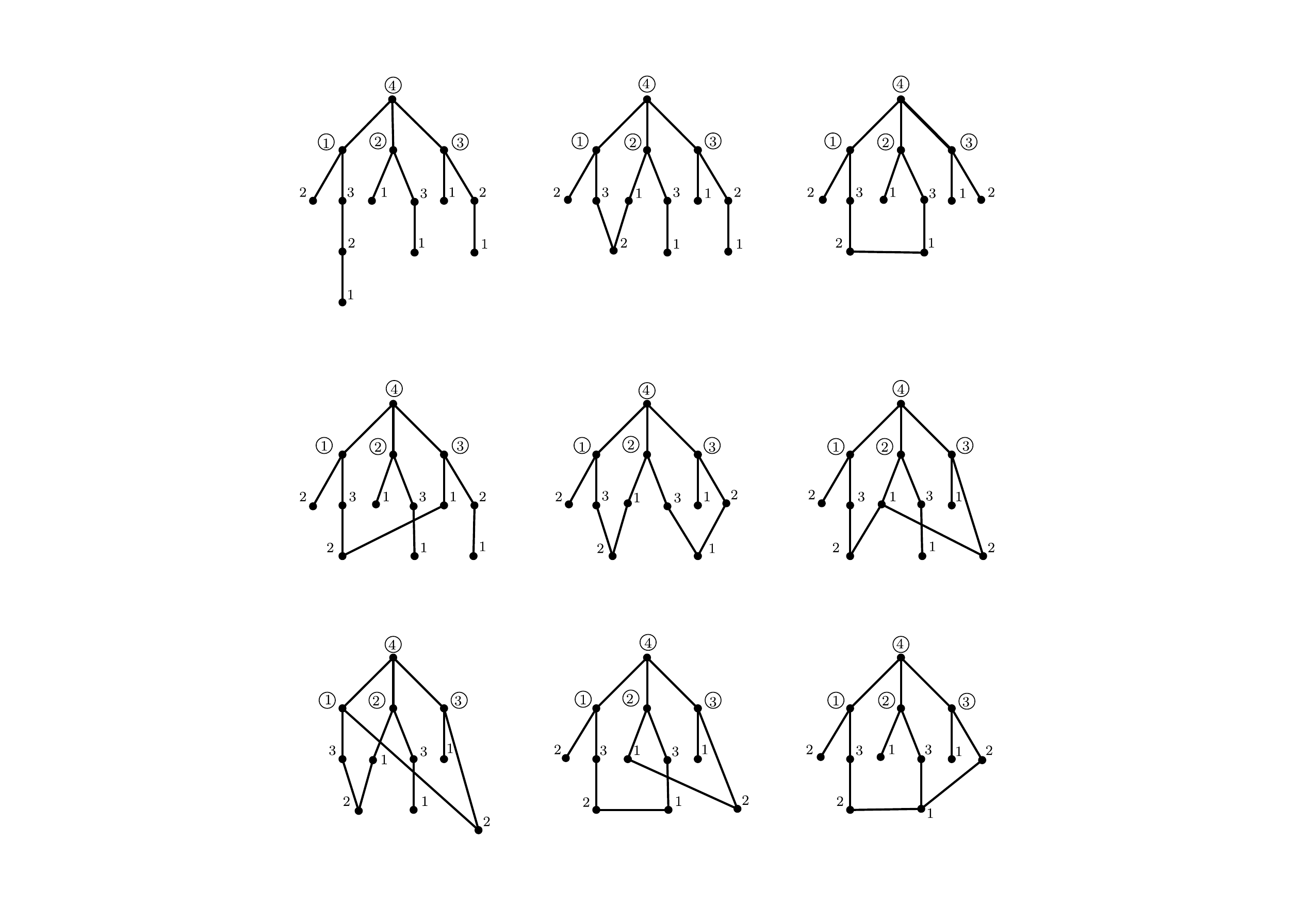}
\caption{The nine $z$-atoms from the whole eighteen triangle-free $z$-atoms with $z$-number four. For each one a $z$-coloring using four colors is illustrated, where the color-dominating vertices are indicated by circles.}\label{1z4}
\end{figure}

\begin{figure}	
	\hspace*{-4cm}	
\includegraphics[width=25cm,height=20cm]{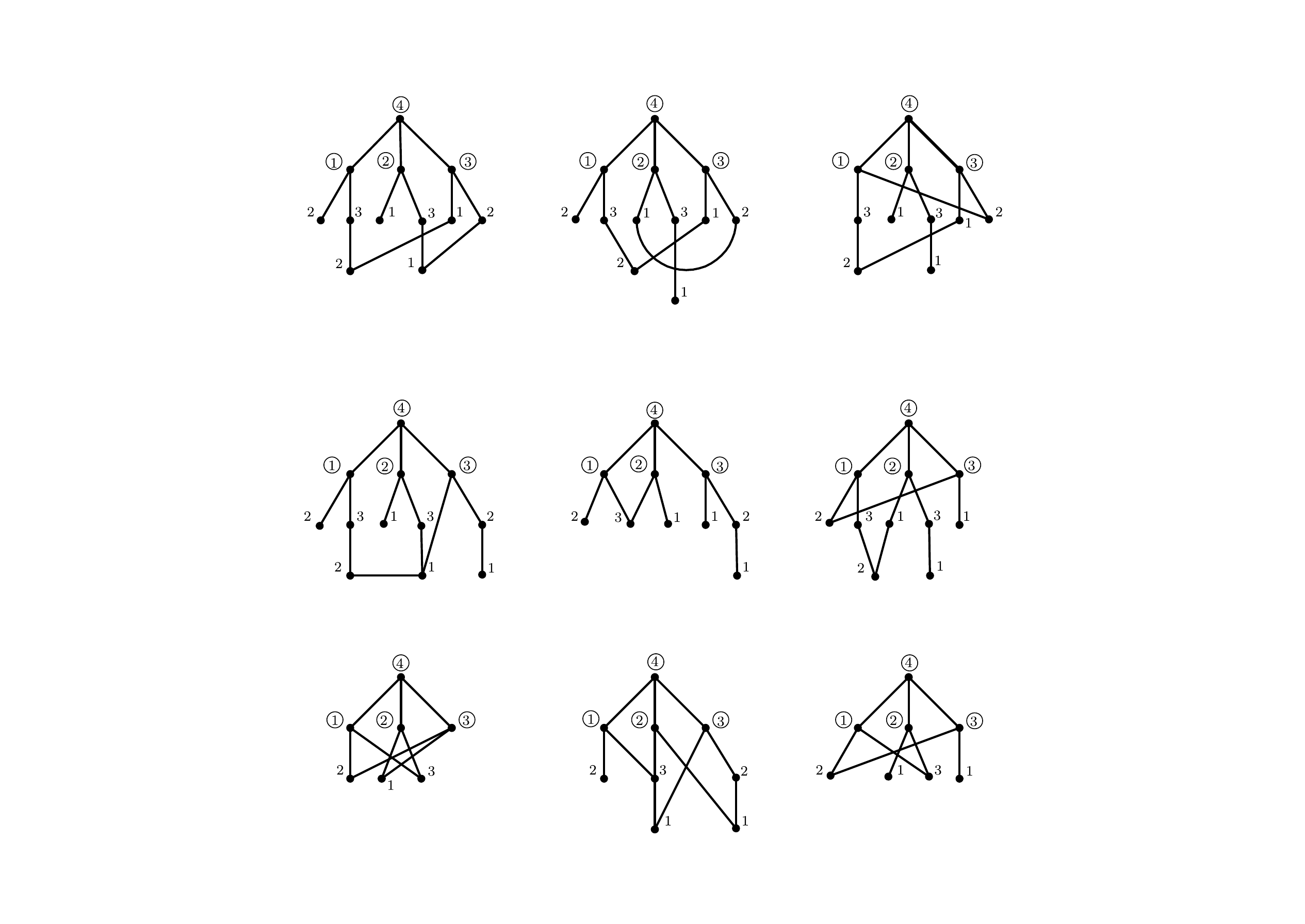}
\caption{The remaining nine $z$-atoms from the whole eighteen triangle-free $z$-atoms with $z$-number four. For each one a $z$-coloring using four colors is illustrated, where the color-dominating vertices are indicated by circles.}\label{2z4}
\end{figure}

\noindent Now we explain the final steps of the construction Phase II. Let $(H,C)$ be any colored graph output from Phase I. In $(H,C)$ the vertex with color $t+1$ is a Grundy vertex. This means that the class $C_{t+1}$ is Grundy class. We operate ${\mathcal{G}}_t$ on $(H,C)$ and obtain a family of colored graphs. Then operate ${\mathcal{G}}_{t-1}$ on each member of this family and obtain a new larger family denoted by ${\mathcal{G}}_{t-1}{\mathcal{G}}_t (H,C)$. Then operate ${\mathcal{G}}_{t-2}$ on the members of the new family and continue this method by applying the operations ${\mathcal{G}}_k$, $k=t-2, t-3, \ldots, 2$. We obtain a final family of colored graphs which can be represented by ${\mathcal{G}}_2 \cdots {\mathcal{G}}_{t-1}{\mathcal{G}}_t (H,C)$. Note that each color class in each member of the new family is Grundy class. The family ${\mathcal{D}}_{t+1}$ is defined as this final family, i.e. $${\mathcal{D}}_{t+1}=\bigcup_{(H,C):(H,C)~is~constructed~in~phase~I} {\mathcal{G}}_2 \cdots {\mathcal{G}}_{t-1}{\mathcal{G}}_t(H,C)$$
\noindent We call each member of ${\mathcal{D}}_{t+1}$ a $z$-atoms with $z$-chromatic number at least $t+1$. Also for each member $G\in {\mathcal{D}}_{t+1}$ the corresponding coloring $C'$ of $G$ is a $z$-coloring of $G$ using $t+1$ colors. This coloring is called the canonic $z$-coloring of $G$. In fact the original vertices of the initial star graph $S_{1,t}$, i.e. $u_1, u_2, \ldots, u_{t+1}$ are color-dominating vertices and by Remark \ref{grundify} the coloring $C'$ of $G$ has Grundy property. Clearly, the complete graphs on one and two vertices are the only $z$-atoms with $z$-number one and two, respectively. It is also easily observed that there are two $z$-atoms with $z$-number three. They are the complete graph $K_3$ and the path on 5 vertices $P_5$. There are too many $z$-atoms with $z$-number four. If we generate only triangle-free ones then we obtain 18 such $z$-atoms, which are depicted in Figures \ref{1z4} and \ref{2z4}, together with a $z$-coloring for each of them consisting of four color-dominating vertices. The largest one is naturally a tree and contains 14 vertices and is illustrated in Figure \ref{1z4}.

\begin{thm}
Let $t\geq 0$ be any fixed integer. Let $L$ be any graph and $z(L)\geq t+1$. Then there exists a $z$-atom $(G,C^{\ast})\in {\mathcal{D}}_{t+1}$ which is embedded in $L$, where $C^{\ast}$ is the canonic coloring of $G$.\label{z-atom}	
\end{thm}

\noindent \begin{proof}
Let $C$ be a $z$-coloring of $L$ using $t+1$ colors. By Proposition \ref{phaseI} there exist $t+1$ vertices $u_1, u_2, \ldots, u_{t+1}$ of colors $1, 2, \ldots, t+1$ in $C$, respectively and a colored graph $(H,C')$ such that for each $i\leq t$, $u_{t+1}$ is adjacent to $u_i$ and $u_i$ is a color-dominating vertex of color $i$. Also started from the star graph consisting of the vertices $u_1, u_2, \ldots, u_{t+1}$, the graph $(H,C')$ is a subgraph of $L$ and one of the graphs generated in Phase I, where the coloring $C'$ is the restriction of the coloring $C$ to the vertices of $H$. Denote the color classes in $H$ by $C_1, C_2, \ldots, C_t, C_{t+1}$. Obviously $C_{t+1}=\{u_{t+1}\}$ is a Grundy class in $H$. Let $D$ be the set of vertices of color $t$ in $H$. Let $w$ be any vertex in $D$. Since $C$ has Grundy property, for each $i<t$ there exists a neighbor of $w$ in $L$ whose color in $C$ is $i$. Let $D_i$ be a minimal subset of vertices in $L$ whose color in $C$ is $i$ and $D_i$ dominates the vertices of $D$. Obviously $D_i$ is partitioned into $D_i\cap V(H)$ and $D_i\setminus V(H)$. Write for simplicity $\tilde{D}_i=D_i\setminus V(H)$. Let $H'$ be a (colored) subset of $G$ induced by $V(H)\cup \tilde{D}_1 \cup \tilde{D}_2 \cup \cdots \cup \tilde{D}_{t-1}$. Interpret the vertices of $\tilde{D}_i$ as the vertices denoted by $w^i_1, \ldots, w^i_{m_i}$ (for some suitable $m_i$) in the operation ${\mathcal{G}}_t$. Also since $D_i=\tilde{D}_i \cup (D_i \cap V(H))$ dominates the vertices of $D\subseteq V(H)$ then there exists a subset of vertices, say tentatively $Q$, in $D$ (this subset $Q$ is in fact denoted by $C_t^i\setminus S_i$ in the operation ${\mathcal{G}}_t$) such that $Q$ is dominated only by $\tilde{D}_i$. Hence the vertices of $Q$ are mapped by a surjective mapping say $g_i$ into $\tilde{D}_i$ (or $\{w^i_1, \ldots, w^i_{m_i}\}$). Denote by $C''$ the coloring of $H'$ obtained by restriction of $C$ to $V(H')$. It follows that one of the graphs constructed in ${\mathcal{G}}_t(H,C')$ is isomorphic to $(H',C'')$. By Remark \ref{grundify} the class of vertices of color $t$ in $(H',C'')$ is Grundy class. Now we repeat the above argument for the colored graph $(H',C'')$ and the color $t-1$ and obtain the next supergraph of $H'$. By continuing this procedure we obtain a colored subgraph of $(L,C)$, say $G$, which is isomorphic to a graph in the family ${\mathcal{G}}_2 \cdots {\mathcal{G}}_{t-1}{\mathcal{G}}_t (H,C')$. Let $C^{\ast}$ be the restriction of $C$ to $V(G)$. Now, the $z$-atom $(G,C^{\ast})$ is embedded in $L$, as desired.
\end{proof}

\noindent Theorem \ref{z-atom} provides a computational tool to prove upper bounds of the form $z(G)\leq t$ for $z$-chromatic number (and so chromatic number) of graphs, where $t$ is any arbitrary and fixed integer. Let $G$ be any graph and $t$ a fixed integer such that no element of ${\mathcal{D}}_t$ is embedded in $G$. Then $z(G)\leq t$. Since ${\mathcal{D}}_t$ is finite, to verify that no element of ${\mathcal{D}}_t$ is embedded in $G$ can be done in a polynomial time steps in terms of $|V(G)|$ by using the following method. It is clear that the largest $z$-atom in ${\mathcal{D}}_t$ is a smallest tree $T$ with $z(T)=t$. It is proved in Proposition \ref{tree} that there exists only one tree $T$ with $z(T)=t$ and with the minimum number of vertices. It is proved that $|V(T)|=(t-3)2^{t-1}+t+2$. Let $f(t)=(t-3)2^{t-1}+t+2$. Let $(H,C)$ be an arbitrary member in ${\mathcal{D}}_t$. The graph $G$ is our input graph. By an exhaustive search in ${|V(G)| \choose |V(H)|}$ steps we can check if $(H,C)$ is imbedded in $G$ or not. Since $|V(H)|\leq f(t)$ then each element of ${\mathcal{D}}_t$ can be checked in $|V(G)|^{f(t)}$ steps. Let $|{\mathcal{D}}_t|=g(t)$. Therefore to verify that no element of ${\mathcal{D}}_t$ is embedded in $G$ needs overall $g(t)|V(G)|^{f(t)}$ time steps. Since $t$ is fixed
then it is polynomial in terms of $|V(G)|$.

\noindent The following result uses the above-mentioned technique but we have to omit some of its lengthy and tedious details.

\begin{thm}
Let $G$ be $(K_3, P_5)$-free graph. Then $z(G)\leq 3$.\label{K3P5}
\end{thm}

\noindent \begin{proof}
Assume on the contrary that $G$ admits a $z$-coloring using 4 colors. Since $G$ is triangle-free then there exists at least one $z$-atom $(H,C)$ from the graphs depicted in Figures \ref{1z4} and \ref{2z4} such that $(H,C)$ is embedded in $G$, where $C$ is the $z$-coloring of $H$ using four colors illustrated in the figures. Applying the following argument for each of these 18 $z$-atoms leads to contradiction. Note that each graph $(H,C)$ in the figures have many induced $P_5$. But $(H,C)$ is embedded in $G$ and $G$ is $P_5$-free. Therefore many extra edges should be added to each $z$-atom $(H,C)$ in order to destroy all existing induced $P_5$. But we are not allowed to add edges between vertices with a same color in $(H,C)$. When we add extra edges to $(H,C)$ satisfying this condition then some new induced $P_5$ is created in the graph. Hence more extra edges are required to be added to the graph. Eventually, this procedure makes a triangle in the graph. This contradicts the fact that $G$ is triangle-free.
\end{proof}

\section{Results concerning trees}

\noindent In this section we first determine the following quantity
$$a_k=\min \{|V(T)|:~T~is~a~tree~with~z(T)=k\}.$$
\noindent We also show that there exists a unique tree $R_k$ such that $z(R_k)=k$ and $|V(R_k)|=a_k$. Then in the light of Theorem \ref{z-atom} it follows that for every tree $T$, $z(T)\geq k$ if and only if $T$ contains a subtree isomorphic to $R_k$. In this situation we can apply the following result of \cite{VR}. Given any two trees $T$ and $R$, to determine whether $R$ is isomorphic to a subtree of $T$ can be solved in polynomial time.

\begin{figure}
\hspace{1cm}
	\includegraphics[width=13cm,height=4.5cm]{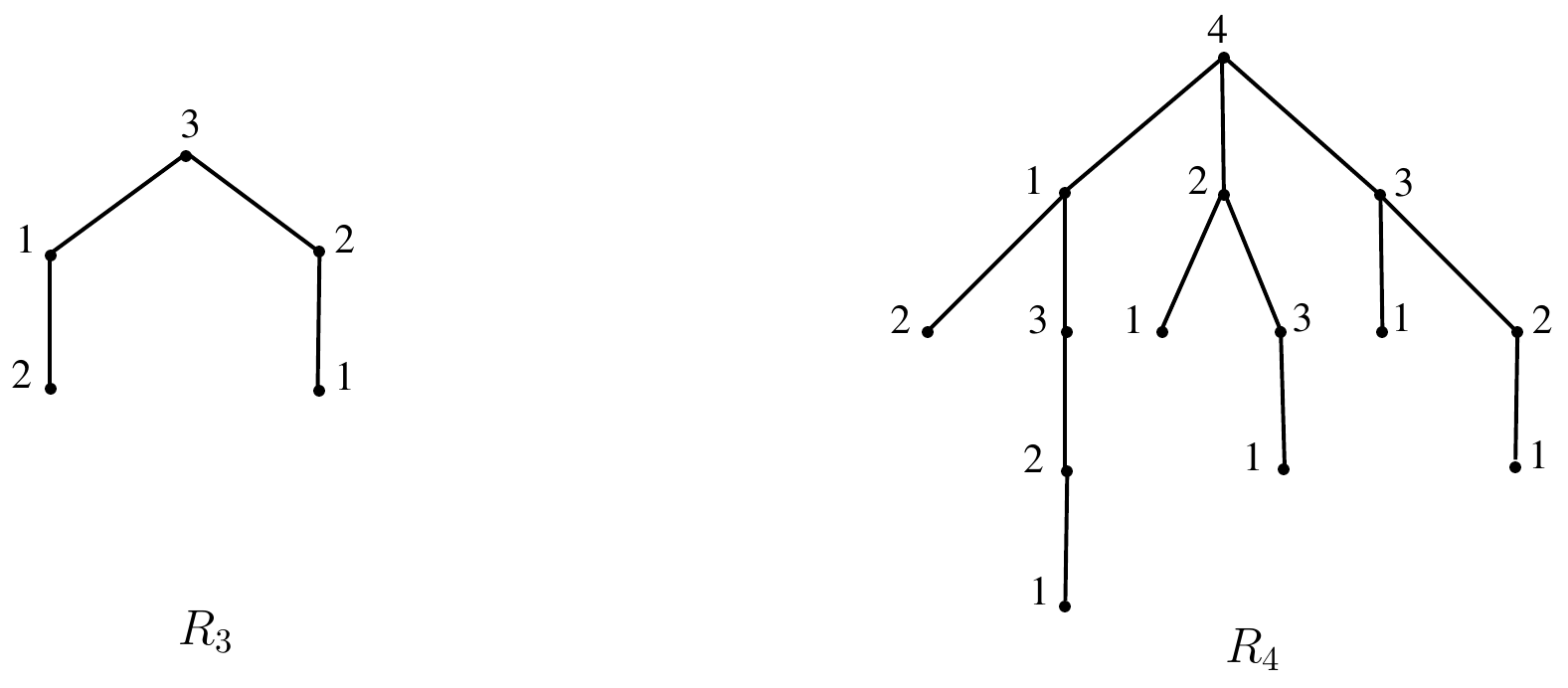}	
	\caption{The trees $R_3$ and $R_4$ with their canonic labelings}\label{treefig}
\end{figure}

\noindent In the following we try to construct a graph $T$ with smallest possible number of vertices satisfying $z(T)=k$ and subject to condition that $T$ is acyclic. Let $C$ be a $z$-coloring with $k$ colors. There should be color-dominating vertices $u_1, \ldots, u_k$ such that $u_j$ is of color $j$ in $C$ and that $u_k$ is adjacent to each $u_j$, $j\not= k$.
Consider $u_k$ as a root and expose the rest of vertices from top to down. Hence, $u_1, \ldots, u_{k-1}$ are the children of $u_k$ and lie in the second or lower level. For each $j\in \{1, \ldots, k-1\}$, $u_j$ is color-dominating vertex of color $j$, so it needs $k-1$ neighbors having colors $1, 2, \ldots, j-1, j+1, \ldots, k-1$. All of such neighbors are distinct because the graph to be constructed is acyclic. Place these neighbors in a third level. But the coloring $C$ has Grundy property, hence the vertices of the third level need suitable new neighbors in the forth level and so on. We denote by $R_k$ the tree constructed according to this procedure. Obviously $R_1$ and $R_2$ are isomorphic the complete graphs $K_1$ and $K_2$, respectively. Observe that $R_3$ is isomorphic to the path on 5 vertices $P_5$. Figure \ref{treefig} depicts $R_3$ and $R_4$ with their corresponding labeling. The tree $R_k$ is constructed so that $z(R_k)\geq k$ and is minimal tree with this property. Also each vertex in the tree has at most one and $k-2$ neighbors in its upper and lower level, respectively. Hence, $\Delta(R_k)=k-1$ and then $z(R_k)=k$. We have $|R_k|=a_k$ by the definition of $a_k$. Note that $a_1=1, a_2=2, a_3=5, a_4=14$. In the following we prove that $a_k=2a_{k-1}+2^{k-1}-k$, for each $k\geq 2$.

\noindent Denote by $S_k$ the subtree of $R_k$ whose root is $u_1$. In the following we make a close connection between $S_k$ and $T_k$, where $T_k$ is the only tree atom whose Grundy number is $k$. Consider $S_k$ and its coloring from $C$ which is a $z$-coloring with $k$ colors. The vertex $u_1$ has color $1$ in this coloring. As we mentioned, $u_1$ has $k-1$ neighbors of colors $2, 3, \ldots, k$ in $C$. For a moment assign color $k$ to $u_1$ and attach a leaf of color $1$ to each neighbor of $u_1$ as well as the vertex $u_1$ itself. Denote the resulting tree by $\tilde{S}_k$. The resulting coloring for $\tilde{S}_k$ is a Grundy coloring for $\tilde{S}_k$ with $k$ colors. Since $R_k$ is minimal then $\tilde{S}_k$ is minimal with respect to Grundy coloring. In other words, $\tilde{S}_k$ is isomorphic to $T_k$. In order to obtain a precise connection, we do the following. In the Grundy coloring of $T_k$ with $k$ colors there exists exactly one vertex of color $k$. Denote this vertex by the very notation $u_1$. The vertex $u_1$ (in $T_k$) has $k-1$ neighbors of colors $1, 2, \ldots, k-1$. Denote these neighbors by $w_1, \ldots, w_{k-1}$. For each $i\geq 2$, $w_i$ has a neighbor say $p_i$ of color 1 in $T_k$. Now remove $u_1, w_1, p_2, \ldots, p_{k-1}$ from $T_k$. Denote the resulting graph by $L_k$. The above argument shows that $L_k$ and $S_k\setminus \{u_1\}$ are isomorphic and the label of each vertex $v$ in $L_k$ is the same as the isomorphic copy of $v$ in $S_k\setminus \{u_1\}$. Note that $|V(L_k)|=|V(T_k)|-k$. It follows that $|V(S_k\setminus \{u_1\})|=2^{k-1}-k$.

\noindent Consider now $U_k= R_k\setminus (S_k\setminus \{u_1\})$. Each vertex in $U_k$ has a neighbor of color 1. See this situation in Figure \ref{treefig}. Remove from $U_k$ these neighbors of color 1 and then decrease the color of each vertex by one. This results in a $z$-coloring with $k-1$ colors for the remaining tree which has $|V(U_k)|/2$ vertices. It follows by the uniqueness of $R_{k-1}$ that the very remaining tree is isomorphic to $R_{k-1}$. Therefore $|V(U_k)|=2|V(R_{k-1})|$. Finally, we obtain that $|V(R_k)|=2|V(R_{k-1})|+2^{k-1}-k$, as desired.

\noindent In order to determine the exact value of $|V(R_k)|=a_k$ we apply the generating function method. Set $f(x)={\sum}_{k=1}^{\infty} a_kx^k$. We have $a_k=2a_{k-1}+2^{k-1}-k$ for each $k\geq 2$ and $a_0=0, a_1=1$. We have,

$$\sum_{k=2}^{\infty} a_kx^k = 2 \sum_{k=2}^{\infty} a_{k-1} x^k + \sum_{k=2}^{\infty} 2^{k-1} x^k - \sum_{k=2}^{\infty} kx^k.$$

\noindent By computations and power series expansions we obtain

$$f(x)= \frac{2x^2}{(1-2x)^2} - \frac{x}{(1-x)^2(1-2x)} + \frac{2x}{(1-2x)}.$$

\noindent By solving $f(x)$ in terms of power series, we obtain $a_k=(k-3)2^{k-1}+k+2$. The following proposition and corollary are immediate.

\begin{prop}
Let $T$ be any tree and $z(T)=k$ for some integer $k$. Then $|V(T)|\geq (k-3)2^{k-1}+k+2$. Moreover, inequality holds only for the tree $R_k$.\label{tree}
\end{prop}

\begin{remark}
The $z$-chromatic number of trees is determined by a polynomial time algorithm.
\end{remark}

\noindent In the following for every graph $H$ denote by $\tilde{H}$ the graph obtained by attaching a leaf to each vertex of $H$. The following result will be used in the next theorem.

\begin{thm}
For any tree $T$, $$\Gamma(T) \leq z(T)^2.$$\label{gammaz}
\end{thm}

\noindent \begin{proof}
Take the trees $R_{k-1}$ and $T_{k-1}$ and consider the trees $\tilde{R}_{k-1}$ and $\tilde{T}_{k-1}$. Let $u$ and $w$ be the roots (with color $k-1$) of $R_{k-1}$ and $T_{k-1}$, respectively. Now connect $\tilde{R}_{k-1}$ and $\tilde{T}_{k-1}$ by adding an edge between $u$ and $w$ and obtain a new tree $R$. By the explanation concerning the construction of $R_k$, note that $R_k$ is a subtree of $R$. In fact the color of $u$ and $w$ will be respectively $k$ and $1$, in the canonic coloring of $R_k$. To prove the theorem it is enough to prove that the tree $T$ with $\Gamma(T)\geq k^2$ contains $R$ as subgraph. Consider a Grundy coloring $C$ of $T$ using $k^2$ color classes $C_1, \ldots, C_{k^2}$. We can embed $\tilde{T}_{k-1}$ in $C_1\cup \cdots \cup C_{k}$ and (by induction on $k$) embed ${R}_{k-1}$ in the next $(k-1)^2$ color classes of $C$. Hence $\tilde{R}_{k-1}$ is embedded in $C_1 \cup C_{k+1} \cup \cdots \cup C_{(k-1)^2+k}$. Since the whole coloring is Grundy then the embedding can be done so that $u$ is adjacent to $w$. Note that $(k-1)^2+k\leq k^2$. This completes the proof.
\end{proof}

\noindent Recall that $T_k$ is the unique smallest tree with $\Gamma(T_k)=k$. The following result compares $z(T_k)$ and $\Gamma(T_k)$.

\begin{thm}
$$\lim_{k\rightarrow \infty} (\Gamma(T_k)-z(T_k))= +\infty.$$
\end{thm}

\noindent \begin{proof}
Set for simplicity $z(T_k)=p_k$. By Proposition \ref{tree} we have
$$2^{k-1}=|T_k|\geq (p_k-3)2^{p_k-1}+p_k+2 > (p_k-3)2^{p_k-1}.$$
\noindent It follows that $2^{k-p_k} > p_k-3$ and $k-p_k > \log (p_k-3)$. Now, if $k-p_k \nrightarrow \infty$ then there exits a number $N$ such that for each $k$,
$k-p_k\leq N$. Hence $\log (p_k-3) < N$ and $p_k< 2^N+3$ for each $k$. But by Theorem \ref{gammaz}, $\sqrt{k} \leq p_k <2^N+3$, a contradiction.
\end{proof}

\section{Further researches}

\noindent Since the $z$-coloring heuristic and $z$-chromatic number are newly defined chromatic concepts, many chromatic and algorithmic problems can be raised for them. Is $z$-chromatic number of graphs an $NP$-complete parameter? As a computational project, it is useful to produce the bank of $z$-atoms, at least for low $z$-numbers. Another research area is to prove upper bound results such as Theorem \ref{K3P5}, using the bank of $z$-atoms. For example, what is the best possible upper bound for the $z$-chromatic number of $(K_4, P_5)$-free graphs? We believe that a competitive coloring heuristic, which we denote by $IZ$, is obtained from the $z$-coloring heuristic. Assume that we applied the $z$-coloring heuristic for a graph $G$ and obtained the color classes $C_1, C_2, \ldots, C_k$. We can repeat the heuristic in the reverse order i.e. from $C_1$ to $C_k$. Also, let $\sigma$ be any random permutation of $\{1, 2, \ldots, k\}$. We can repeat the heuristic by scanning the classes according to the order $C_{\sigma(1)}, C_{\sigma(2)}, \ldots, C_{\sigma(k)}$. Call this iterated form of the heuristic, iterated $z$-coloring heuristic (shortly $IZ$). Since every $z$-coloring has Grundy property then $IZ$ is better than the iterated greedy $IG$ heuristic of Culberson \cite{C, CL}.

\section{Acknowledgment}

\noindent The author thank anonymous referees for their useful comments.

\end{document}